%% file: main.tex
\documentclass[10pt,conference, dvipsnames]{IEEEtran}
\usepackage{cite}
\usepackage{amsmath,amssymb,amsfonts}
\usepackage{graphicx}
\usepackage{textcomp}
\usepackage[hyphens]{url}

\usepackage{amsmath,amsfonts}
\usepackage{graphicx}
\usepackage{textcomp}
\usepackage[dvipsnames]{xcolor}  
\usepackage{xspace}
\usepackage{nicefrac}
\usepackage[flushleft]{threeparttable}
\usepackage{booktabs}
\usepackage{multirow}
\usepackage{makecell}
\usepackage{algorithm}
\usepackage[noend]{algpseudocode}
\usepackage{subcaption}
\usepackage[per-mode=symbol, range-units=single, mode=text]{siunitx}
\usepackage{pifont}
\usepackage{listings}
\usepackage{inconsolata}
\usepackage[colorlinks, linkcolor=black, citecolor=blue, urlcolor=black]{hyperref}
\usepackage[capitalise, nameinlink]{cleveref} 

\newcommand{\tablefontsize}{\footnotesize}

\newcommand{\design}{KAPLA\xspace}

\newcommand{\dir}[1]{\texttt{#1}}
\newcommand{\tensorcolor}{RoyalBlue}
\newcommand{\stackcolor}{ForestGreen}
\newcommand{\updatecolor}{BurntOrange}
\newcommand{\tensor}[1]{\dir{\textcolor{\tensorcolor}{\textbf{tensor}}{#1}}}
\newcommand{\stack}[1]{\dir{\textcolor{\stackcolor}{\textbf{stack}}{#1}}}
\newcommand{\update}[1]{\dir{\textcolor{\updatecolor}{\textbf{update}}{#1}}}
\newcommand{\tensorname}[1]{\tensor{\{{#1}\}}}

\newcommand{\bytes}[1]{\SI{#1}{B}}
\newcommand{\kb}[1]{\SI{#1}{\kilo B}}
\newcommand{\mb}[1]{\SI{#1}{\mega B}}

\newcommand{\revised}[1]{{\color{Black} #1}}
\newcommand{\deleted}[1]{{}}
\newcommand{\shrink}[2][]{#1}
\newcommand{\newshrink}[2][]{#1}

\def\BibTeX{{\rm B\kern-.05em{\sc i\kern-.025em b}\kern-.08em
    T\kern-.1667em\lower.7ex\hbox{E}\kern-.125emX}}

\title{{\design{}: Pragmatic Representation and Fast Solving
of Scalable NN Accelerator Dataflow}}

\author{Zhiyao Li\\
Tsinghua University\\
{\tt\small lizhiyao19@mails.tsinghua.edu.cn}
\and
Mingyu Gao\\
Tsinghua University\\
{\tt\small gaomy@tsinghua.edu.cn}
}

\begin{document}
\maketitle
\thispagestyle{plain}
\pagestyle{plain}

\begin{abstract}
\input{0_abstract}
\end{abstract}

\input{1_introduction}
\input{2_background_motivation}
\input{3_taxonomy_directives}
\input{4_solver}
\input{5_methodology}

\input{6_evaluation}
\shrink{\input{7_related_work}}
\input{8_conclusion}

\bibliographystyle{IEEEtranS}
\bibliography{refs}

\end{document}

%% file: 0_abstract.tex
Dataflow scheduling is of vital importance to neural network (NN) accelerators. Recent scalable NN accelerators support a rich set of advanced dataflow techniques.
The problems of comprehensively representing and quickly finding optimized dataflow schemes thus become significantly more complicated and challenging.
In this work, we first propose comprehensive and pragmatic dataflow representations for temporal and spatial scheduling on scalable multi-node NN architectures. 
An informal hierarchical taxonomy highlights the tight coupling across different levels of the dataflow space as the major difficulty for fast design exploration. 
A set of formal tensor-centric directives accurately express various inter-layer and intra-layer schemes, and allow for quickly determining their validity and efficiency. 
We then build a generic, optimized, and fast dataflow solver, KAPLA. It makes use of the pragmatic directives to explore the design space with effective validity check and efficiency estimation. 
KAPLA decouples the upper inter-layer level for fast pruning, and solves the lower intra-layer schemes with a novel bottom-up cost descending method. 
KAPLA achieves within only 2.2\% and 7.7\% energy overheads on the result dataflow for training and inference, respectively, compared to the exhaustively searched optimal schemes. It also outperforms random and machine-learning-based approaches, with more optimized results and orders of magnitude faster search speedup.


%% file: 1_introduction.tex
\section{Introduction}

As Dennard scaling ends, modern computer systems have been embracing domain-specific acceleration besides general-purpose processors~\cite{jh_dp_turing_lecture_18}. An efficient accelerator design has to not only involve novel hardware architectures trimmed to the domain characteristics, but also incorporate software-level mapping and scheduling frameworks that ensure optimized utilization of the 
precious resources. As an important domain, neural networks (NNs) have recently seen a large number of academic and industrial acceleration efforts on improving both the hardware architectures~\cite{diannao_asplos14, eie_isca16, tpu_isca17, scaledeep_isca17, simba_micro19} and the software dataflow paradigms~\cite{eyeriss_isca16, tangram_asplos19, timeloop_ispass19, maestro_micro19, interstellar_asplos20}.

As modern NNs keep evolving to contain increasingly larger and more complicated structures~\cite{googlenet_cvpr15, resnet_cvpr16, lstm_gnmt_nips14}, we need to also scale to larger accelerators with much more on-chip compute and storage resources for better performance and efficiency~\cite{tpu_isca17, scaledeep_isca17, simba_micro19}. The abundant resources create more opportunities for new dataflow patterns and enable a rich set of dataflow techniques.
We can schedule a single layer individually~\cite{eyeriss_isca16, cnvlutin_isca16, tpu_isca17}, or pipeline multiple layers simultaneously~\cite{cnn_pipeline_fpl16, pipelayer_hpca17, scaledeep_isca17};
\shrink{we can force all model weights to stay on-chip~\cite{brainwave_dnn_fpga_isca18}, or forward intermediate data between layers in different granularities~\cite{tangram_asplos19};}
we can spatially parallelize the computation in different dimensions~\cite{tetris_asplos17, nn_para_gpu_sysml19, timeloop_ispass19}, or temporally block the data to balance the data reuse~\cite{blocking_cnn_arxiv16, timeloop_ispass19}.

However, at the same time, it is also becoming more challenging to achieve high resource utilization on these scalable accelerators with optimized dataflow scheduling. There has not been a comprehensive and unified framework that can systematically and completely capture the huge design space of existing NN dataflow schemes, nor is there an effective methodology to quickly identify optimized schedules for a given workload on a specific hardware. Such a design space is highly complicated and challenging, as it covers from multiple chips/chiplets/tiles to thousands of individual processing elements, and involves several levels of memory hierarchy. Most previous work only targeted specific subproblems of the complete dataflow paradigm with different assumptions, producing insights and results that might be hard to generalize. Moreover, existing scheduling algorithms to find optimized dataflow schemes are insufficient in terms of both search speed and resulted efficiency. They use either naive exhaustive search running for unacceptably long time~\cite{blocking_cnn_arxiv16, timeloop_ispass19}, or machine-learning-based heuristics that can only apply to small, focused subspaces~\cite{autotvm_neurips18, chameleon_rl_iclr20}.

To overcome these issues, we present \emph{comprehensive and pragmatic dataflow representations} for dense NNs, which cover a rich set of existing \emph{temporal} and \emph{spatial} dataflow schemes at both \emph{inter-layer} and \emph{intra-layer} levels, and support both \emph{training} and \emph{inference} workloads. We first illustrate an informal \emph{hierarchical dataflow taxonomy}. This taxonomy not only exhibits the large design space at each inter- and intra-layer level, but, more importantly, also reveals the tight coupling across different levels, which is the root reason that hinders us from conducting quick design space exploration without sacrificing efficiency. Then, we introduce a set of formal \emph{tensor-centric dataflow directives} that improve over prior work~\cite{maestro_micro19, interstellar_asplos20}. By treating the tensors stored throughout the accelerator memory hierarchy as first-class citizens, we can directly inspect both the data sizes occupying a buffer and the data access volumes across buffers, which are crucial for determining whether a dataflow scheme is \emph{valid} and/or \emph{efficient}. The directives are thus pragmatic to dataflow solvers, and work well on various hardware architecture templates.

We then build a \emph{generic, optimized, and fast dataflow solver}, \design{}, which could effectively explore the complicated design space with inter-layer and intra-layer, temporal and spatial dataflow schemes, on both large multi-node accelerators and small edge devices.
\design{} leverages two key techniques to effectively handle \emph{validity check} (test if a scheme satisfies all buffering/parallelism constraints) and \emph{efficiency estimation} (estimate the performance/energy of a scheme), the two main tasks in dataflow solvers. 
First, \design{} effectively \emph{decouples the design space}, enabling fast upper-level inter-layer pruning with negligible loss in optimality. 
Second, \design{} \emph{adopts a novel bottom-up cost descending way} to efficiently solve intra-layer schemes, under a uniform description for multi-level memory hierarchies and supporting advanced data sharing options such as systolic~\cite{tpu_isca17} and buffer sharing~\cite{tangram_asplos19}. 
\design{} makes use of our \emph{pragmatic tensor-centric directives} to better capture \emph{spatial resource utilization} and \emph{temporal data movements}, for convenient performance and efficiency evaluation.

Evaluated with a wide range of modern NNs on scalable multi-node accelerators, \design{} achieves within just 2.2\% and 7.7\% energy overheads on average compared to the exhaustively searched optimal, for training and inference, respectively. \design{} can complete such high-quality dataflow scheduling up to two orders of magnitude faster than prior approaches, and typically within \emph{a few minutes} even for large NN training. This is a significant step towards real-time interactive compilation of optimized dataflow on scalable accelerators. We also demonstrate \design{} is robust to different hardware configurations, including small edge accelerators, and both Eyeriss-like and TPU-like architecture templates.

%% file: 2_background_motivation.tex
\section{Background and Motivations}
\subsection{Neural Networks}

A neural network (NN) usually consists of tens to hundreds of \emph{layers}, which can be represented by directed acyclic graphs (DAGs). In \emph{inference} tasks, each input is fed forward through the layers following the DAG topology. In \emph{training}, the original DAGs are extended with more layers for error propagation and weight updates.

In this work, we focus on the most compute-intensive convolutional (CONV) and fully-connected (FC, a.k.a., matrix multiplication) layers. In a CONV layer, each channel of the output feature maps (fmaps) is accumulated across the convolutions between all input fmap channels and a set of filters (a.k.a., weights). Plus batching, all tensors have four dimensions, as our notations in \cref{table:dim_notation}. In FC layers each fmap is simply a single neuron. The backward CONV/FC layers can be modeled similarly to the forward layers with different data layouts and computations~\cite{pipelayer_hpca17, accpar_hpca20}.

\begin{table}
    \centering\tablefontsize
    \caption{Notation for tensor dimensions.}
    \label{table:dim_notation}
    \begin{threeparttable}
    \begin{tabular}{c}
        \toprule
        Input tensor \\
        \dir{N}, \dir{C}, \dir{Xi}, \dir{Yi} \\
        \midrule
        Output tensor \\
        \dir{N}, \dir{K}, \dir{Xo}, \dir{Yo} \\
        \midrule
        Weight tensor \\
        \dir{K}, \dir{C}, \dir{R}, \dir{S} \\
        \bottomrule
    \end{tabular}
    \end{threeparttable}
    \hspace{1em}
    \begin{threeparttable}
    \begin{tabular}{cl}
        \toprule
        \dir{N} & Batch \revised{size} \\
        \dir{C} & Input fmap channel \revised{size} \\
        \dir{K} & Output fmap channel \revised{size} \\
        \dir{Xi}, \dir{Yi} & Input fmap width \& height \\
        \dir{Xo}, \dir{Yo} & Output fmap width \& height \\
        \dir{R}, \dir{S} & Filter weight width \& height \\
        \bottomrule
    \end{tabular}
    \end{threeparttable}
\end{table}

\subsection{Scalable NN Accelerators and Dataflow
}
\label{subsec:baseline}

\begin{figure}
  \centering
  \includegraphics[width=0.85\columnwidth]{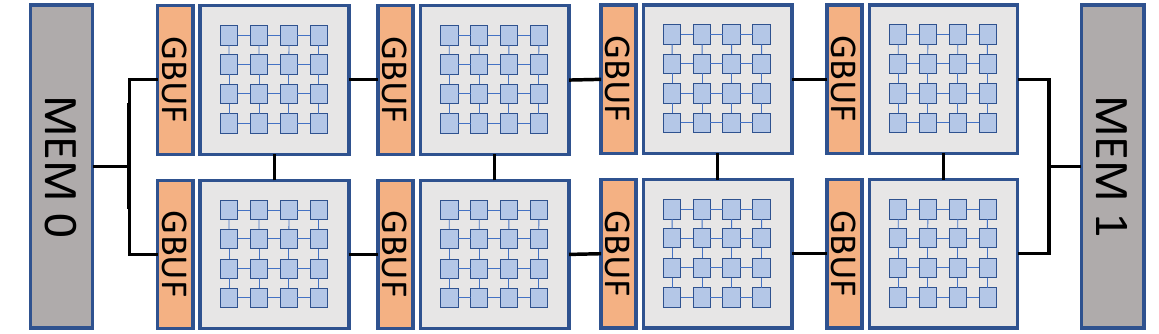}
  \caption{A multi-node NN accelerator. Each node has an array of PEs, a register file per PE, and a global SRAM buffer.}
  \label{fig:multi_node_arch}
\end{figure}

The importance and high performance requirements of NNs have made them an emerging domain for specialized hardware acceleration. Many existing NN accelerators are derived from a common spatial architecture consisted of an array of processing elements (PEs) and a hierarchy of on-chip buffers, e.g., global buffers (GBUF) and register files (REGF)~\cite{neuflow_cvprw11, diannao_asplos14, shidiannao_isca15, eyeriss_isscc16, cambricon_isca16, flexflow_hpca17, tpu_isca17}. As more complex and deeper NNs require continuously increasing amount of computation, recent NN accelerators have to \emph{scale} their compute capabilities. A common way to do so is to adopt a parallel, multi-node architecture on a single chip or across multiple chips/chiplets~\cite{dadiannao_micro14, neurocube_isca16, tetris_asplos17, scaledeep_isca17, tangram_asplos19, simba_micro19}.
In this work, we focus on a representative architecture shown in \cref{fig:multi_node_arch}. It interconnects a group of nodes and multiple off-chip memories using a network-on-chip (NoC) or an inter-chip/chiplet network.

With the abundant compute and buffer resources offered by modern NN accelerators, it is not trivial to achieve the best utilization of such powerful hardware. Finding the best \emph{dataflow schemes} to schedule various NN layers thus becomes a critical task. 
Prior research has proposed a wide range of optimized dataflow schemes, such as systolic or stationary styles of spatial data transfers in PE arrays~\cite{eyeriss_isca16, tpu_isca17}, loop transformation heuristics to improve data reuse in on-chip buffers~\cite{blocking_cnn_arxiv16, timeloop_ispass19}, workload partitioning and coordination across multiple nodes and chips~\cite{cnn_fpga_util_isca17, tetris_asplos17, brainwave_dnn_fpga_isca18, hypar_hpca19, accpar_hpca20}, pipelined processing across layers~\cite{scaledeep_isca17, tangram_asplos19, cnn_pipeline_fpl16}, as well as dataflow specialized for sparse NNs~\cite{scnn_isca17}.
Our multi-node NN accelerator in \cref{fig:multi_node_arch} supports a rich set of dataflow.
\shrink{
At the top level, an arbitrarily large NN can be divided into multiple \emph{segments} of layers. Each segment can fit in the accelerator and is executed separately, one after the other. All layers in a segment execute simultaneously on different nodes in a \emph{pipelined} fashion and with \emph{fine-grained data forwarding}~\cite{scaledeep_isca17, tangram_asplos19}. We also support \emph{weight pinning}~\cite{brainwave_dnn_fpga_isca18} if applicable. Each layer in the segment can use one or multiple nodes, with flexible, \emph{hybrid parallelization} schemes that partition the tensors along batch, output channel, and/or fmap width/height dimensions~\cite{tetris_asplos17, nn_para_gpu_sysml19, hypar_hpca19, accpar_hpca20}. To alleviate buffer capacity pressure, shared tensors can be stored across multiple nodes with the \emph{buffer sharing} dataflow~\cite{tangram_asplos19}. Within each node, the PE array adopts the \emph{row stationary} scheme from Eyeriss~\cite{eyeriss_isca16}, or the \emph{systolic} style from TPU~\cite{tpu_isca17}. And \emph{optimized loop transformation} schemes are used to minimize the off-node data transfers~\cite{blocking_cnn_arxiv16, timeloop_ispass19}.}
A complete and systematic description can be found in \cref{subsec:taxonomy}.

\subsection{Motivations and Challenges}
\label{subsec:challenges}

Although various dataflow optimizations for scalable NN accelerators greatly improve performance and efficiency, they exhibit an extraordinary large design space, which is continuously expanding with newly emerging optimizations. To quickly understand and effectively exploit the fast evolving hardware and dataflow techniques, we should (a) introduce \emph{comprehensive} and \emph{pragmatic} representations that describe the complete optimization space; (b) develop \emph{generic}, \emph{optimized}, and \emph{fast} dataflow scheduling solutions that effectively explore the numerous design options.

\textbf{Representing dataflow options.}
With diverse hardware scales (single-node, multi-node, multi-chip/chiplet), various dataflow techniques (as in \cref{subsec:taxonomy}), and numerous NN structures, there still lacks a \emph{comprehensive} way to cover the complete dataflow design space.
Each prior work uses its own taxonomy to describe a restricted design space, either limited to a single layer~\cite{eyeriss_isca16, timeloop_ispass19, maestro_micro19, interstellar_asplos20}, or solely focusing on coarser granularities of parallelism~\cite{hypar_hpca19, accpar_hpca20, tetris_asplos17, tangram_asplos19}. As different dataflow levels are tightly coupled and affected by each other, a uniform framework is needed to find global optimal solutions.
More importantly, besides just describing all dataflow solutions, the representation should also be \emph{pragmatic} to facilitate efficient dataflow search and reflect key information such as data traffic patterns and buffer capacity utilization. 

There has been a large body of previous work on understanding and representing NN dataflow. Nested loop structures are the most generic and widely used approach, e.g., by Yang et al.~\cite{blocking_cnn_arxiv16} and Timeloop~\cite{timeloop_ispass19}. MAESTRO~\cite{maestro_micro19} transformed the loops into temporal and spatial mapping directives. NN-Baton~\cite{nnbaton_isca21} similarly extended the temporal and spatial directives to package/chiplet-level parallelism. Interstellar~\cite{interstellar_asplos20} borrowed directives from the Halide language~\cite{halide_pldi13}. TENET~\cite{tenet_isca21} proposed more flexible, relation-centric notations to characterize arbitrary affine loop transformations.

Unfortunately, these representations are not friendly to dataflow solvers.
Using generic nested loop structures~\cite{timeloop_ispass19, blocking_cnn_arxiv16} requires complex and recursive procedures that are hard to reason about. For example, deriving tensor sizes and access traffic at a buffer needs to analyze relevant blocked loops at \emph{all} higher levels in a deep memory hierarchy. Changing even one blocked loop factor would affect the sizes, access counts, and parallelization degrees of all related tensors, because these statistics are \emph{heavily entangled} in the nested loop representation. 
The temporal/spatial directives~\cite{maestro_micro19, nnbaton_isca21} and the Halide/Interstellar scheduling primitives~\cite{interstellar_asplos20} are more pragmatic, but still manipulate individual loop dimensions rather than tensors and suffer from the same issues. 
TENET is able to explore flexible and irregular dataflow, but sacrificing solver speed due to the much larger design space and the higher complexity to evaluate each scheme~\cite{tenet_isca21}. 
Actually, none of these representations was proposed \emph{together with} a non-trivial search algorithm, which implies that they did not include pragmatic exploration as the primary goal.

To determine the optimality of a dataflow scheme, the solver needs to investigate: (1) the \emph{spatial resource utilization} of the parallel compute units (nodes, PEs) and the on-chip buffers (GBUF, REGF), and (2) the \emph{temporal data movements} between these buffer levels through interconnects.
Such information mainly reflects how the \emph{tensors} are \emph{buffered} in and \emph{accessed} through the memory hierarchy, and would be best captured by a representation which treats tensors as first-class citizens.

\textbf{Solving optimized dataflow.}
A good dataflow solver should find \emph{optimized} schemes for a \emph{generic} set of hardware and software configurations, and also do so sufficiently \emph{fast}. The last requirement may seem unnecessary, as the search is a one-time cost and the result schemes are deployed to long-run tasks. We argue it is not the case any more.

First, after enlarged with the inter-layer partitioning and pipelining schemes~\cite{tetris_asplos17, tangram_asplos19, hypar_hpca19, accpar_hpca20, cnn_pipeline_fpl16}, the complete design space now requires hours or even \emph{days} to explore for a large-scale NN on a multi-node architecture using naive exhaustive search (see \cref{subsec:results}).
Second, advanced hardware acceleration and software optimizations are now able to reduce the training time of large NNs down to only a few hours or even minutes~\cite{large_batch_resnet_arxiv17, large_batch_dp_arxiv2019}, making the relative cost of dataflow scheduling considerably more significant.
Third, \revised{in many cases dataflow scheduling is not a one-time cost. Design space exploration of custom accelerators needs to evaluates the target NN models on many hardware configurations to find the best performance and energy points.}
The widely adopted network architecture search (NAS) techniques explore a large number of NN structure candidates~\cite{efficientNAS_parameter_sharing_arxiv18, NAS_rl_arxiv16}. Many, if not all, layers must be re-scheduled due to different topologies and/or layer dimensions, in order to find the optimized dataflow for fast training speed, and also for accurate inference performance estimation~\cite{mnasnet_cvpr19, fbnet_cvpr19}. 
Finally, with the emerging Machine-Learning-as-a-Service (MLaaS), if the clients are allowed to supply their own models, fast compilation and dataflow scheduling at the cloud side would be desired.

Unfortunately, existing techniques fail to achieve the above three requirements simultaneously. A large body of prior work adopts brute-force exhaustive search~\cite{blocking_cnn_arxiv16, eyeriss_isca16, tangram_asplos19}. While generic and optimized, such search is quite \emph{slow} with large NNs and multi-node accelerators. Alternatively, heuristic-based algorithms speed up the exploration by either conducting a random search~\cite{timeloop_ispass19}, or applying machine-learning (ML) methods~\cite{autotvm_neurips18, chameleon_rl_iclr20, mindmappings_asplos21, magnet_iccad19}. These heuristics are \emph{less generic} in the complete design space. For example, AutoTVM~\cite{autotvm_neurips18} learned a tree-based model to guide the optimizations only on individual layers, and Mind Mappings~\cite{mindmappings_asplos21} used an MLP to approximate the limited design space of only intra-layer and single-node architectures. ML is also difficult to rule out ``invalid'' schedules on specialized accelerators, such as those whose blocked data fail to fit in on-chip buffers~\cite{timeloop_ispass19}. \shrink[Different from CPU/GPU caches]{While implicitly managed CPU/GPU caches can tolerate these cases with degraded performance}, accelerator buffers require explicit and precise data orchestration with hard constraints~\cite{pencil_pact2015, cosa_isca21}.
CoSA~\cite{cosa_isca21} modeled dataflow scheduling as mixed-integer programming (MIP) and used a commercial MIP solver. Although it only takes a few seconds to schedule a single layer, the MIP formulation is unlikely to scale to larger inter-layer spaces. \revised{This is because the complexity of MIP grows exponentially with the number of variables, which would be significantly enlarged by a factor equal to the number of layers (10s to 100s) when considering both inter- and intra-layer parallelism.}

Furthermore, most prior efforts just focus on their own sub-problems, e.g., either limited to a single layer~\cite{eyeriss_isca16, timeloop_ispass19, maestro_micro19, interstellar_asplos20, cosa_isca21, mindmappings_asplos21, tenet_isca21}, or solely focusing on coarser granularities of parallelism~\cite{hypar_hpca19, accpar_hpca20, tetris_asplos17, tangram_asplos19, nnbaton_isca21}. 
\revised{For example, device placement scheduling on multiple GPUs~\cite{device_placement_rl_icml2017, device_placement_rl_iclr2018} focuses on high-level inter-layer parallelism but lacks intra-layer exploration.}
The insights from these solvers may be incompatible or even conflicting due to different assumptions for the rest of the design space. Simply combining them will likely lead to suboptimal overall results. For instance, \shrink{the data buffer bypass heuristic in TETRIS~\cite{tetris_asplos17} is not compatible with layer pipelining that requires fully buffering fmaps on-chip~\cite{pipelayer_hpca17, scaledeep_isca17}.
As another example, }though all the following three studies simultaneously processed multiple layers, Li et al.~\cite{cnn_pipeline_fpl16} scheduled the entire NN with a single batch, Gao et al.~\cite{tangram_asplos19} scheduled a subset of the NN, while Shen et al.~\cite{cnn_fpga_util_isca17} scheduled different batches. It is therefore difficult to put them together to aggregate their benefits.

To overcome these challenges, we first present a pragmatic representation that comprehensively and hierarchically describes the dataflow space with tensor-centric directives (\cref{sec:representation}). Built upon this representation, we propose a generic, optimized, and fast dataflow solver \design{} (\cref{sec:solver}).

%% file: 3_taxonomy_directives.tex
\section{Dataflow Representations}
\label{sec:representation}

\subsection{A Hierarchical Dataflow Taxonomy}
\label{subsec:taxonomy}

Previous NN dataflow taxonomies limit their scopes to a single layer and mostly on single-node accelerators~\cite{eyeriss_isca16, maestro_micro19, interstellar_asplos20}. We target more complex multi-node architectures, and hierarchically compose both \emph{temporal} and \emph{spatial} schemes at \emph{inter-layer} and \emph{intra-layer} levels. \cref{table:taxonomy} summarizes previous proposals into our taxonomy.

\begin{figure}
  \centering
  \includegraphics[width=\columnwidth]{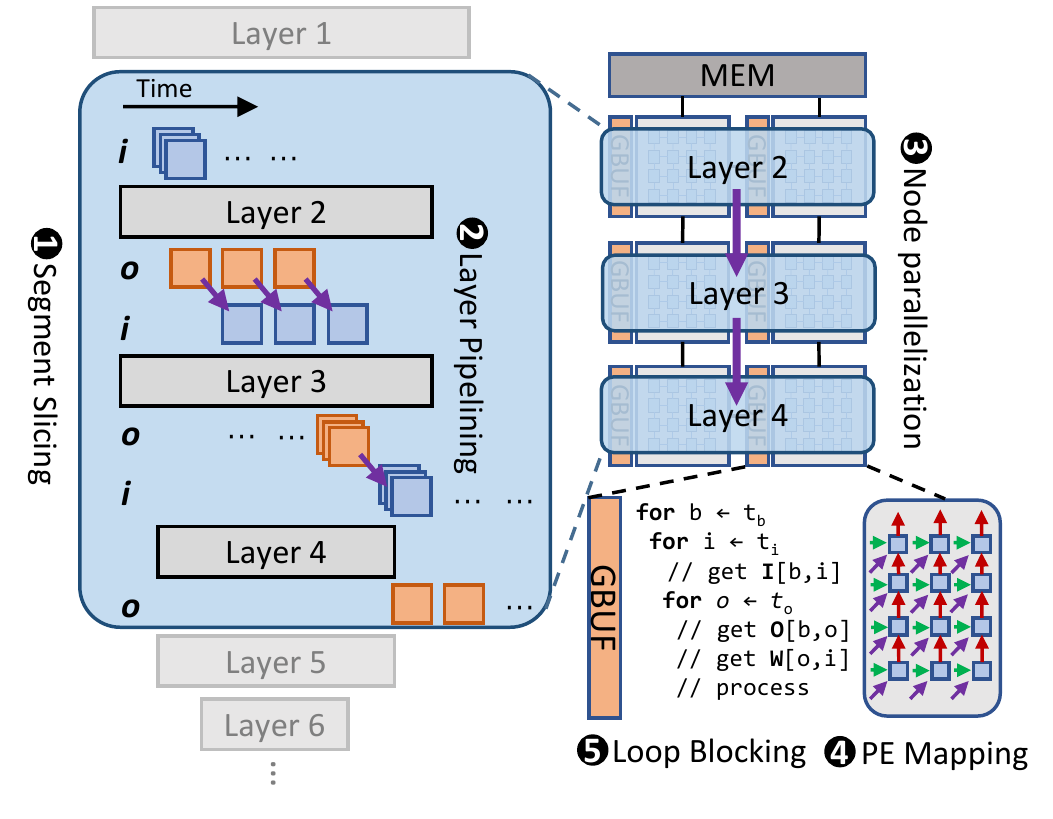}
  \caption{The hierarchical NN dataflow taxonomy.}
  \label{fig:taxonomy}
\end{figure}

\begin{table}
    \centering\tablefontsize
    \caption{Previous work summarized using our hierarchical dataflow taxonomy.}
    \label{table:taxonomy}
    \begin{threeparttable}
    \begin{tabular}{ccl}
        \toprule
        \multirow{2}{*}{\shortstack{Inter-\\layer}}
        & Temporal & Segment slicing~\cite{tangram_asplos19, cnn_fpga_util_isca17, device_placement_rl_icml2017, device_placement_rl_iclr2018} \\
        & Spatial & Layer pipelining~\cite{tangram_asplos19, cnn_pipeline_fpl16, pipelayer_hpca17, neocpu_atc19, scaledeep_isca17} \\
        \midrule
        \multirow{3}{*}{\shortstack{Intra-\\layer}}
        & Spatial & Node parallelization~\cite{neurocube_isca16, tetris_asplos17, hypar_hpca19, accpar_hpca20, cosa_isca21, nnbaton_isca21} \\
        & Spatial & PE mapping~\cite{eyeriss_isca16, tpu_isca17, flexflow_hpca17, shidiannao_isca15, maeri_nn_itcn_asplos18, timeloop_ispass19, tenet_isca21} \\
        & Temporal & Loop blocking~\cite{blocking_cnn_arxiv16, tetris_asplos17, timeloop_ispass19, autotvm_neurips18, chameleon_rl_iclr20, cosa_isca21, mindmappings_asplos21} \\ 
        \bottomrule
    \end{tabular}
    \end{threeparttable}
\end{table}

\textbf{Inter-layer.}
Inter-layer dataflow can be categorized into \emph{temporal} and \emph{spatial} scheduling across multiple layers. First, \emph{segment slicing} temporally schedules different groups of consecutive layers (called \emph{segments}) from a large NN on the same accelerator~\cite{tangram_asplos19, cnn_fpga_util_isca17}. Different segments time-share the hardware to utilize the limited resources. All layers within a segment, such as Layers 2, 3, and 4 in \cref{fig:taxonomy} \ding{182}, execute together. Then, \emph{layer pipelining} spatially schedules the layers in a segment onto different nodes of the multi-node accelerator. The intermediate fmap tensors are stored in the on-chip buffers and directly forwarded through the interconnects between nodes in a pipelined manner~\cite{cnn_pipeline_fpl16, pipelayer_hpca17}.
\revised{Layer fusion~\cite{fused_layer_cnn_micro16} could be modeled as special layer types as they require specific architectural support on spatial accelerators.}

The choice of segment lengths exhibits an important tradeoff. With shorter segments, each layer can use more compute and buffer resources and executes more efficiently. In contrast, longer segments eliminate more intermediate data off-chip transfers. On the other hand, layer pipelining can use different choices of granularity and timing for the intermediate data forwarding. For example, in \cref{fig:taxonomy} \ding{183}, one fmap is forwarded from Layer 2 to Layer 3 each time, while all three fmaps from Layer 3 are forwarded together to Layer 4. It has been demonstrated that finer-grained data forwarding reduces the latency as well as the intermediate buffer occupation~\cite{tangram_asplos19}. But the granularities are subject to several constraints. Essentially, adjacent layers (e.g., Layers 2 and 3) must have matched access patterns to the same intermediate tensors, so that the tensors can be consumed as soon as produced.

Between the two inter-layer dataflow categories, the design space of layer pipelining is largely shaped by the segment slicing decisions. In turn, the latency and buffer requirements of layer pipelining also affect the feasibility of the segments.

\textbf{Intra-layer.}
Intra-layer dataflow schemes can be described using loop transformations on the layer's nested loops~\cite{blocking_cnn_arxiv16, timeloop_ispass19, interstellar_asplos20}. In general, at each level of the accelerator's memory hierarchy, loop unrolling results in \emph{spatial} parallelization, while loop blocking and reordering optimize \emph{temporal} data reuse. Our discussion below assumes two levels of on-chip buffers (PE registers and node buffers) in a multi-node architecture, but also generalizes to deeper levels.

At the node level, \emph{node parallelization} spatially parallelizes the layer across nodes. In \cref{fig:taxonomy} \ding{184}, we partition the computation of each layer onto two nodes.  State-of-the-art schemes typically partition workloads in a hybrid way along multiple dimensions~\cite{tetris_asplos17, nn_para_gpu_sysml19, hypar_hpca19}, including batch \dir{N}, channels \dir{C} and \dir{K}, and 2D fmap \dir{Xo}, \dir{Yo}, \dir{Xi}, \dir{Yi}. When a certain dimension is selected, the tensors that contain this dimension are partitioned into smaller subtensors, each of which becomes easier to fit in the buffer of one node to reduce off-chip accesses. In contrast, the other shared tensors remain the same, and need to be replicated in different nodes, taking even more buffer capacity. To avoid such data duplication, the buffer sharing optimization~\cite{tangram_asplos19} stores a single copy of the shared tensor across multiple buffers, and rotates each other's share to allow every node to have a chance of local access at different time.

Similarly at the PE level, \emph{PE mapping} spatially distributes the computations onto the PE array. Some tensors can be partitioned and stay stationary within each PE~\cite{eyeriss_isca16, flexflow_hpca17}, while the others are transferred to multiple PEs through multicasting~\cite{eyeriss_v2_jetcas19}, or across neighbor PEs in a systolic flow~\cite{tpu_isca17} (\cref{fig:taxonomy} \ding{185}). We notice that, the PE-level data multicasting and the node-level buffer data duplication are similar, as both transfer and replicate the shared data in multiple units; the PE-level systolic flow and the node-level buffer sharing are similar, as they avoid data replication by only storing a subset of data in each unit and gradually fetching new data from neighbors. These similarities could help describe different memory levels in a uniform way in \cref{subsec:hardware}.

Finally, even after spatial distribution, in most cases the tensors are still too large to fit entirely in the on-chip buffers. \emph{Loop blocking and reordering} have been extensively studied to improve data reuse across the memory hierarchy, by dividing the tensors into smaller blocks and finding the best order to fetch them~\cite{blocking_cnn_arxiv16, timeloop_ispass19} (\cref{fig:taxonomy} \ding{186}). These schemes are difficult to optimize, especially considering the large number of blocking and reordering choices over the many nested loop dimensions. It is therefore common to use exhaustive, random, or heuristic search methods~\cite{blocking_cnn_arxiv16, eyeriss_isca16, timeloop_ispass19}.

\textbf{Summary.}
\revised{The above taxonomy indicates tight coupling between all the inter- and intra-layer levels.}
The overall intra-layer space is constrained by the chosen inter-layer schemes, including the number of nodes assigned to the layer, and the desired top-level access granularity and pattern to match with the adjacent layers. Between the intra-layer levels themselves, as all the above transformations are applied on the same nested loops, their design spaces are naturally coupled. For example, spatially parallelizing over more nodes will reduce the subtensor sizes in one node, and result in less needed temporal loop blocking.
\revised{Hence we must holistically consider the full, tightly coupled design space, which complicates scheduling search.}

\subsection{Tensor-Centric Dataflow Directives}
\label{subsec:directives}

The informal taxonomy in \cref{subsec:taxonomy} has comprehensively illustrated the dataflow options in a qualitative way. We now propose a formal and quantitative representation with a set of \emph{tensor-centric directives}, which will be utilized by our dataflow solver in \cref{sec:solver}.

The philosophy behind our tensor-centric directives is to construct the dataflow from the inside out along the memory hierarchy, capturing \emph{spatial resource utilization} and \emph{temporal data movements}. These statistics are particularly critical to quickly determine if a dataflow scheme is valid (all tensors fit in their buffers), efficient (minimum data accesses), and with high performance (maximally parallelized). We first declare the tensors to be allocated and processed on a single PE. Then we extend to the entire PE array by leveraging homogeneity, spatially replicating or sharding tensors with different offsets across PEs. Finally we characterize temporal data movements for each computation iteration. We do these recursively for each level (PEs, nodes, etc.). Similar to Halide~\cite{halide_pldi13}, we omit the compute operations in the directives.

\shrink{The directives are intended to be used by an automatic tool (e.g., our solver in \cref{sec:solver}) to explore the design space given NN and hardware configurations. Programmers do not write the directives, but the tool uses them to rewrite dataflow descriptions and eventually identifies an optimized scheme.}

\textbf{Directive definitions.}
\tensor{(dim=size, \textrm{\ldots}[, shr])} declares a multi-dimensional (sub)tensor allocated in the buffer at a certain level of the memory hierarchy, with the given \dir{size} along each \dir{dim}. The optional sharing factor \dir{shr} is used when the tensor is stored across multiple buffers in a shared manner and each buffer only holds $1/\text{\dir{shr}}$ of the data, e.g., with buffer sharing~\cite{tangram_asplos19}. Multiple \tensor{} definitions at the same level can be in any order. Tensors are the basic objects other directives operate on.

\tensor{} declarations are organized in a two-level name scope, across NN layers and across memory hierarchy levels (\cref{lst:directive_example}). Tensors are properly named to capture data dependencies across layers in an NN. Within each layer, tensors with the same name but at different memory levels belong to the same full tensor, and tensors in faster levels are subsets of tensors in slower levels. Effectively, the sizes of the slower level set the spatially and temporally mapping bounds for the sizes of the fast level. Such a naming scheme relieves us from explicitly specifying the detailed tensor \emph{layout} (e.g., how it is tiled) in each buffer, since this can be automatically inferred from the dimension sizes of the tensors in neighboring level buffers that have the same name. For example, a tensor with \dir{C=3} and \dir{C=6} in two neighboring levels means that it should be tiled by 2.

\begin{figure}
  \centering
  \includegraphics[width=0.9\columnwidth]{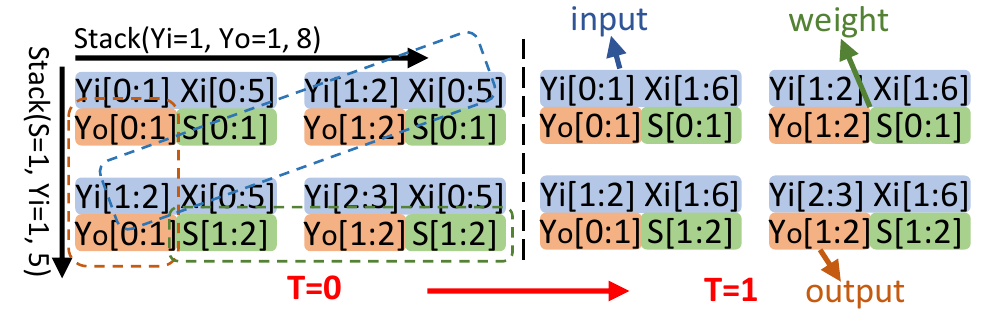}
  \caption{Illustration for \dir{REGF} \stack{} and \update{} directives.}
  \label{fig:stack_illustration}
\end{figure}

\stack{(dim+=shift, \textrm{\ldots}, repl)} represents the spatial parallelization where \dir{repl} PEs or nodes exist at a certain level of the memory hierarchy, each with a local buffer that stores a copy of \emph{all} tensors declared at \emph{this} level with the same defined sizes, but at potentially distinct offsets differing by \dir{shift}s along these \dir{dim}s. Basically, these tensors are \emph{stacked} together across the \dir{repl} buffers; or in another word, a tensor is sharded across them. Note that \dir{shift}s do not need to match the tensor \dir{size}s. A smaller \dir{shift} partially overlaps the tensors, while a larger \dir{shift} distributes non-contiguous ranges to the buffers. 
This allows us to freely specify fine-grained interleaving or coarse-grained partitioning.
If no \dir{shift}s are given, the same tensor is replicated.
\stack{}s in the same level are recursively applied in their declared order. This allows us to specify hybrid and complex spatial parallelism. For example, two \stack{}s in \cref{lst:directive_example:stack-1,lst:directive_example:stack-2} at the \dir{REGF} level of \cref{lst:directive_example} indicate a 2D PE array mapping, as illustrated in \cref{fig:stack_illustration}.

\update{(dim+=step, \textrm{\ldots})} represents the ordered and nested temporal iterations in individual buffers, which characterizes the iterative data movements across memory levels.
In each iteration, \emph{all} tensors across all the buffers at \emph{this} level are updated at the same pace, incremented by \dir{step}s along the \dir{dim}s simultaneously. The new data evict the old data from the buffers. Again, \dir{step}s do not need to equal to tensor \dir{size}s, allowing for flexible data updates such as overlapped sliding windows and non-contiguous strides.

\begin{lstlisting}[
float,
caption={Directive examples for CONV and DWCONV layers, with row-stationary PE mapping~\cite{eyeriss_isca16}, output + batch hybrid node parallelization~\cite{tetris_asplos17}, and layer pipelining~\cite{tangram_asplos19}.},
label=lst:directive_example,
emph={tensor}, emphstyle={\color{\tensorcolor}\bfseries},
emph={[2]stack}, emphstyle={[2]\color{\stackcolor}\bfseries},
emph={[3]update}, emphstyle={[3]\color{\updatecolor}\bfseries},
emph={[4]CONV,DWCONV,REGF,GBUF}, emphstyle={[4]\bfseries},
]
CONV:
 REGF:
  tensor{0}(N=1, C=2, Xi=5, Yi=1)
  tensor{w1}(C=2, K=3, R=5, S=1)
  tensor{1}(N=1, K=3, Xo=1, Yo=1)
  stack(Yi+=1, Yo+=1, 8)  % PE columns (*\label{lst:directive_example:stack-1}*)
  stack(S+=1, Yi+=1, 5)   % PE rows    (*\label{lst:directive_example:stack-2}*)
  update(Xi+=1, Xo+=1)    % 1D conv    (*\label{lst:directive_example:update-1d-conv}*)
  update(Yi+=8, Yo+=8)    % folding    (*\label{lst:directive_example:update-folding}*)
  update(N+=1)
  update(C+=2)
  update(K+=3)
 GBUF:
  tensor{0}(N=4, C=4, Xi=19, Yi=19, shr=4)          (*\label{lst:directive_example:tensor-bufshr}*)
  tensor{w1}(C=4, K=6, R=5, S=5)
  tensor{1}(N=4, K=6, Xo=15, Yo=15)
  stack(K+=6, 4)    % output node parallel    (*\label{lst:directive_example:stack-outp}*)
  stack(N+=4, 16)   % batch node parallel
  update(C+=4)
  update(K+=24)
  update(N+=64)
DWCONV:
 REGF:
  % ...
 GBUF:
  % DWCONV input is the same as CONV output
  tensor{1}(N=4, C=4, Xi=9, Yi=15)     (*\label{lst:directive_example:tensor-intermediate}*)
  tensor{w2}(C=4, R=3, S=3)
  tensor{2}(N=4, C=4, Xo=4, Yo=7)
  stack(C+=4, 6)    % channel node parallel
  stack(N+=4, 16)   % batch node parallel
  stack(Xo+=4, 2)   % output width node parallel
  update(Yo+=7)
  update(C+=24)
  update(N+=64)
\end{lstlisting}

\begin{figure}
  \centering
  \includegraphics[width=0.8\columnwidth]{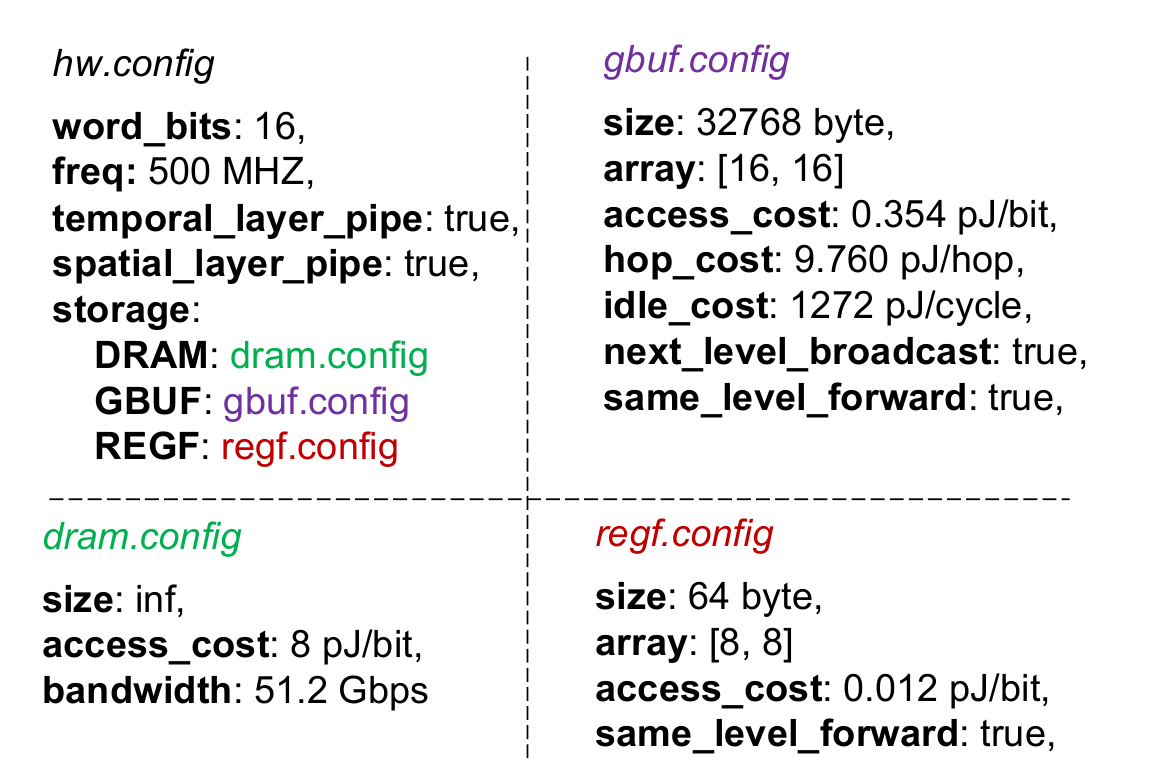}
  \vspace{-1em}
  \caption{An example hardware configuration.}
  \label{fig:hw_config_example}
\end{figure}

\textbf{An example.}
\cref{lst:directive_example} illustrates a dataflow scheme for two adjacent CONV and depthwise CONV~\cite{mobilenets_arxiv17} layers, described by the tensor-centric directives. The scheme is produced on a multi-node accelerator configuration as \cref{fig:hw_config_example}.

The \dir{CONV:REGF} level represents the row-stationary 2D PE mapping~\cite{eyeriss_isca16}. First, three \tensor{}s are allocated in each PE that does a 1D convolution between one row of the filter weights (\dir{R=5,S=1}) and the corresponding range of one input row (\dir{Xi=5,Yi=1}) to calculate one output element (\dir{Xo=Yo=1}). Then the first \update{} in \cref{lst:directive_example:update-1d-conv} slides over the entire fmap row by processing a new element along \dir{Xi} and \dir{Xo} each cycle, as in \cref{fig:stack_illustration}. Data from multiple channels (\dir{C,K}) are stored in the registers to improve reuse.
The two \stack{}s indicate how data are distributed across the PE array as in \cref{fig:stack_illustration}. The first \stack{} increments \dir{Yo}, so each PE array column corresponds to a different row of the output tensor. If the fmap tensors have more rows than the number of array columns, the \update{} in \cref{lst:directive_example:update-folding} temporally folds the data onto the same physical array. Similarly, the second \stack{} increments \dir{S}, resulting in different rows of the filter weight tensor across the PE array rows. And \dir{Yi} is incremented in both, making rows of the input tensor distributed diagonally. Finally, the last three \update{}s represent the nested loop order.

The \dir{CONV:GBUF} level similarly defines three tensors that will be stored in the node buffer. Here we spatially map the layer onto a region of $4 \times 16$ nodes, by using hybrid output and batch parallelization (\dir{K} and \dir{N})~\cite{tetris_asplos17}. We enable sharing the input tensors across four node \dir{GBUF}s with buffer sharing dataflow~\cite{tangram_asplos19}, which is determined by the \dir{repl} of \stack{} in \cref{lst:directive_example:stack-outp} for output parallelization.

The \dir{DWCONV} section shows how the directives support non-CONV/FC layers, e.g., a depthwise CONV~\cite{mobilenets_arxiv17}. The key difference from the previous \dir{CONV} is \dir{C} and \dir{K} are the same dimension now, so \tensorname{w2} only has \dir{C,R,S}, and the output \tensorname{2} uses the same \dir{C=4} as the input \tensorname{1}. This example also uses a more complex parallelization scheme across $12 \times 16$ nodes, where channel \dir{C}, batch \dir{N}, and output width \dir{Xo} are unrolled by 6, 16, and 2, respectively.

Finally, we notice that the input tensor of \dir{DWCONV} is the same as the output tensor of \dir{CONV}, i.e., \tensorname{1}, indicating inter-layer data dependencies. Their sizes at the \dir{GBUF} level are the same, e.g., \dir{K=6} stacked by \dir{4} vs. \dir{C=4} stacked by \dir{6}. Their top-level loop blocking schemes also match, with \update{(K+=24)} vs. \update{(C+=24)}, and both \update{(N+=64)}. Consequently, they are produced and consumed in the same way by the two layers, forming a fine-grained forwarding pipeline~\cite{tangram_asplos19}.

\textbf{Calculating resource utilization and data movement statistics (data sizes, parallelism, and access counts).}
The size of a tensor is simply the product over all its dimensions. The \stack{} directives directly tell us the parallelism at each level. Combining the corresponding \update{} and \tensor{} directives allows us to easily calculate the amount of data accesses. Take \dir{CONV} \tensorname{1} in \cref{lst:directive_example} as an example. Accessing one output row from \dir{GBUF} (\dir{Xo=15}) requires 15 non-overlapped \update{}s at \dir{REGF} (\dir{Xo+=1}). Since the tensor size at \dir{REGF} is 3, we need $3 \times 15 = 45$ data accesses from \dir{GBUF} to \dir{REGF}. These statistics are useful to our solver in \cref{sec:solver}.

\textbf{Advantages.}
First, by treating tensors as first-class citizens using \tensor{}s, we can directly inspect the blocked data sizes allocated in every buffer in each memory level, and quickly discover any buffer capacity underutilization or overflow. 
By further explicitly describing the changes of the tensors over time with the combination of corresponding \update{} and \tensor{} directives, we can also easily calculate the amount of data transfers across the memory hierarchy. 
These results are of great importance to a dataflow solver, but require complex algorithms in previous frameworks like MAESTRO~\cite{maestro_micro19} and Interstellar~\cite{interstellar_asplos20} (\cref{subsec:challenges}). 

Second, our directives provide the same expressibility as generic nested loops at the intra-layer level, but are more expressive to cover the entire dataflow space. 
In particular, we can represent various inter-layer pipelining schemes on multi-node architectures, using matched \tensor{} names for data dependencies and matched \update{}s for transfer granularities. These are difficult with MAESTRO~\cite{maestro_micro19} or Interstellar~\cite{interstellar_asplos20}, especially for complicated DAGs. 

Finally, our directive space is hierarchical by design, and supports \emph{any number of memory levels and nested parallelism}, as well as \emph{other NN layers} beyond CONV and FC.
We believe tensors are a more effective representation. It is the tensors that are buffered and accessed, determining the dataflow efficiency and greatly simplifying the dataflow solver design.

\subsection{Generic Hardware Configuration Template}
\label{subsec:hardware}

Our directives work with a general hardware architecture template, similar to the multi-node accelerator in \cref{fig:multi_node_arch}. \cref{fig:hw_config_example} shows an example specification for the hardware configurations.
Each memory level has its capacity, bandwidth, and access cost. The numbers of PEs ($8 \times 8$) and nodes ($16 \times 16$) are given through the array sizes of their corresponding levels, \dir{REGF} and \dir{GBUF}. The global options \dir{temporal\_layer\_pipe} and \dir{spatial\_layer\_pipe} enable temporal and spatial inter-layer dataflow optimizations, respectively. 

Data transfers across the memory hierarchy can usually happen in two ways. In simple cases with buses or trees, the data can be fetched from the next level and unicast or multicast to one or multiple buffers at this level (\emph{next-level} transfers). Alternatively, the data can be transferred from a neighbor buffer at the same level if already presenting there (\emph{same-level} transfers). Systolic arrays~\cite{tpu_isca17} and buffer sharing~\cite{tangram_asplos19} use the latter approach. We require the hardware configurations to distinguish these two options for each memory level. If same-level transfers are enabled, our \update{} directives will automatically detect the data in the nearest neighbor buffers and serve the overlapped ranges from there. This also applies to the tensors pipelined across layers, to handle the on-chip data forwarding between node buffers.

Most hardware architectures require specific dataflow across the on-chip PEs~\cite{eyeriss_isca16, shidiannao_isca15, tpu_isca17}. Some recent designs support more flexible mapping schemes~\cite{flexflow_hpca17, maeri_nn_itcn_asplos18}. In such cases, the lowest-level \dir{REGF} dataflow scheme should be either fully fixed or constrained by certain patterns~\cite{interstellar_asplos20}. These can be conveniently and accurately specified using our directives.

%% file: 4_solver.tex
\section{\design{} Dataflow Solver}
\label{sec:solver}

In this section, we present a generic, optimized, and fast dataflow solver, \design{}, that outperforms existing random search and ML-based heuristics. It is able to capture the hierarchical inter- and intra-layer design space on multi-node architectures.
\shrink{The directives simplify validity check and efficiency estimation, allowing \design{} to run very fast. \design{} supports typical inference and training schemes by basing the underlying dataflow on universal DAG topologies and generic layer types.}
\design{} is currently implemented in Python, and will be open-sourced. 

\subsection{High-Level Ideas}

Optimized \emph{and} fast exploration in the hierarchical dataflow space is challenging, because of not only the large number of choices in each of the inter-layer and intra-layer levels, but also the tight coupling between them. The upper-level inter-layer schemes shape the lower-level intra-layer design spaces. And in turn, only until the detailed intra-layer schemes are settled, can we accurately tell whether an inter-layer scheme is efficient or even valid.

Dataflow solvers must effectively handle: (1) \emph{validity check}, i.e., how to quickly determine whether a scheme satisfies all constraints, such as buffer capacities and parallelism limits. (2) \emph{efficiency estimation}, i.e., how to accurately estimate the performance/energy of a scheme, in order to identify potentially optimized candidates. \design{} leverages two key novel techniques for these two tasks, as summarized in \cref{table:solver}.

\begin{table}
    \centering\tablefontsize
    \caption{Key techniques in \design{}.}
    \label{table:solver}
    \begin{threeparttable}
    \begin{tabular}{ccc}
        \toprule
        & {\bf Validity check} & {\bf Efficiency estimation} \\
        \midrule
        {\bf Inter-layer} & \makecell{Conservative \\pruning} & \makecell{Prioritization based on \\estimated cost} \\
        \midrule
        {\bf Intra-layer} & \makecell{Bottom-up tensor \\construction} & \makecell{Cost descending \\search} \\
        \bottomrule
    \end{tabular}
    \end{threeparttable}
\end{table}

First, \emph{\design{} effectively decouples the inter-layer and intra-layer spaces, enabling fast upper-level inter-layer pruning without sacrificing optimality} (\cref{subsec:solver-inter-layer}).
Specifically, \design{} supports fast and effective validity check and efficiency estimation \emph{only} using upper-level inter-layer schemes. Many invalid or suboptimal inter-layer schemes can be quickly skipped, without wasting time on solving their detailed lower-level intra-layer schemes. \design{} thus focuses on a small set of potentially more optimized candidates.

Second, \emph{\design{} adopts bottom-up cost descending to efficiently solve intra-layer schemes, under a uniform description for multi-level memory hierarchies with advanced data sharing} (\cref{subsec:solver-intra-layer}). 
Such a solution is fast and optimized. 
By working bottom-up through the memory hierarchy from inner to outer levels, \design{} guarantees the subtensors in each local buffer are always valid, eliminating expensive buffer capacity checks in traditional top-down approaches that repetitively try many loop dimension factorizations. It also effectively guides the search by accurately estimating efficiency.

\begin{figure}
  \centering
  \includegraphics[width=0.85\columnwidth]{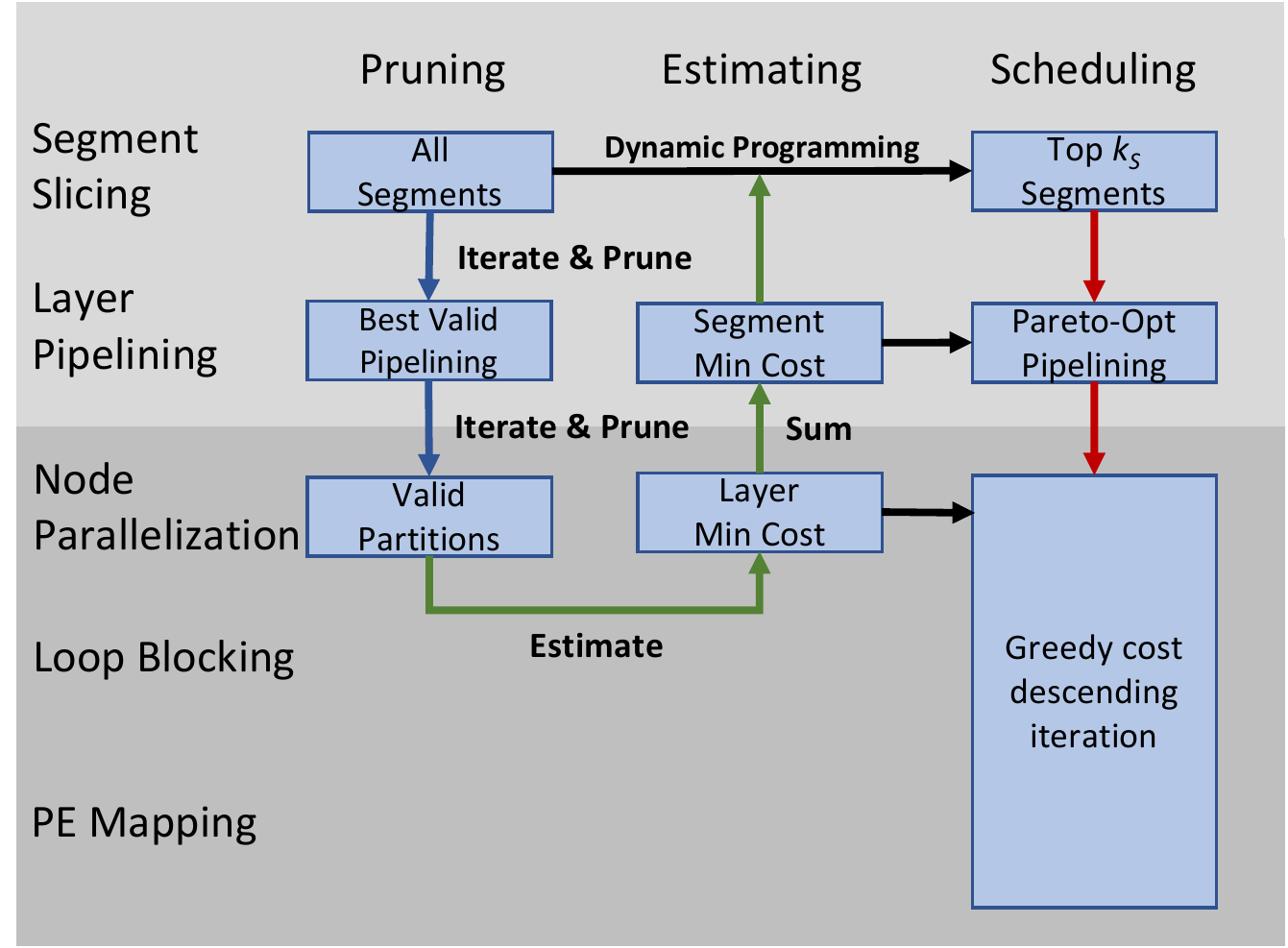}
  \caption{The overall workflow of KAPLA. The blue, green, and red arrow chains represent the pruning, estimating, and scheduling processes, respectively.}
  \label{fig:kapla_workflow}
\end{figure}

\cref{fig:kapla_workflow} illustrates the overall workflow of \design{}. First, we iterate over all possible dataflow schemes for the inter-layer levels, and quickly check their validity by approximately and conservatively calculating the required computation/buffer resources without looking at other levels. This allows effective pruning. We then prioritize the remaining “potentially valid” schemes by their estimated cost with the cost model described above. The top ones are selected for further intra-layer exploration by the greedy cost descending algorithm.

To implement these novel search strategies, \emph{\design{} is implemented upon our pragmatic tensor-centric directives} in \cref{subsec:directives}.
These directives comprehensively cover the dataflow design space, and naturally support the key techniques in \design mentioned above by better capturing \emph{spatial resource utilization} and \emph{temporal data movements} to facilitate validity check and efficiency estimation.
The generic hardware template also allows \design to support advanced and complex data sharing options such as multicast~\cite{eyeriss_v2_jetcas19}, systolic~\cite{tpu_isca17}, and buffer sharing~\cite{tangram_asplos19}, by uniformly modeling the same-level and next-level data transfers (\cref{subsec:hardware}).

\textbf{Cost model.}
\design{} models both energy and performance as simple functions of \emph{resource utilization} (on PEs and buffers) and \emph{data access counts} (on all buffers).
\newshrink{
The energy cost is modeled as $\text{energy} = \sum_{b \in \text{buffers}}{\text{access}_{b} \times \text{cost}_{b}} + \text{hop}_\text{NoC} \times \text{cost}_\text{NoC} + \text{num}_\text{MAC} \times \text{cost}_\text{MAC} + \text{cost}_\text{static}$, which is the weighted sum of the data traffic across bus and the memory hierarchy (DRAM, SRAM, register files), the hop count over NoC, the MAC operations, and the static energy.
}
The latency is estimated with a roofline model composed of the memory hierarchy access latency, the interconnect (bus, NoC) latency, and the MAC operation latency\newshrink{, as $\text{latency} = \max{\{\text{latency}_\text{mem}, \text{latency}_\text{NoC}, \text{latency}_\text{MAC}\}}$}.
\newshrink{
The detailed parameters are derived from actual hardware and modeling tools.
We find these approaches already result in good accuracy (\cref{subsec:sensitivity}). More complicated cost models,
as long as they are also based on these two sets of low-level statistics, can be easily plugged in with future work.
}

\subsection{Inter-Layer Pruning and Prioritization}
\label{subsec:solver-inter-layer}

As elaborated in \cref{subsec:taxonomy}, inter-layer scheduling determines the temporal and spatial resource allocation for a layer segment, through segment slicing and layer pipelining. For validity check, \design uses \emph{conservative pruning} to quickly shrink the feasible design space. For efficiency estimation, it relies on a \emph{fast cost estimation} using \emph{dynamic-programming-based prioritization}, to select high-potential schemes for detailed consideration in the intra-layer phase.

\textbf{Conservative validity pruning.}
\design{} conducts fast and conservative pruning at the inter-layer level, by directly estimating the minimum required buffer capacity of a layer at each memory level, using the best knowledge it has at this point. If the buffer capacity is even smaller than this minimum requirement, the layer is guaranteed to not fit with any possible intra-layer scheme. Therefore such pruning is fast (not exploring detailed intra-layer schemes) and conservative (never rejecting valid inter-layer schemes). Nevertheless, the pruning is quite effective in practice, able to eliminate over 90\% candidates and result in 10$\times$ search speedups (\cref{table:pruning_ablation_study}).

More specifically, with the data buffering constraints for each pipelined layer in a given segment, \design{} ignores the intra-layer spatial and temporal mapping details, and calculates whether the data of this layer can fit in the aggregated buffer capacity across all nodes allocated to it. This check is fast as we only explore pipelining granularities at the outermost buffer level. It is also conservative because indivisible partitioning along a dimension may result in fragmentation, and unpartitioned tensors are replicated in multiple nodes, both increasing actual required capacities. Fortunately, data dimension sizes are typically much larger than the numbers of nodes, and optimizations like buffer sharing~\cite{tangram_asplos19} can alleviate data duplication. So false positive cases (actually invalid but judged as valid) are uncommon, leading to high pruning rates.

\textbf{Fast cost estimation.}
In order to prioritize potentially more optimized schemes, \design{} assigns a cost to each inter-layer scheme of every segment. 
The key insight here is to always approximate to the optimistic cases if there is insufficient information, so the estimated cost would be a (relatively tight) lower bound. When only used to \emph{prioritize} candidates, a lower bound closely reflects the scheme's potential. And our prioritization method is able to tolerate errors (see below).

To show concrete examples, recall our cost model is based on resource utilization and access activities.
While DRAM and inter-node access counts are determined by the dataflow at the top buffer levels, the node buffer accesses are actually based on the inner levels between PEs and nodes. Additionally exploring these levels will enlarge the design space and slow down the search. So instead we use the minimum access count that \emph{may} be achievable. 
For resource utilization, we again assume that the layer could use \emph{all} the PEs across all the nodes assigned to it, temporarily ignoring any internal fragmentation.
And for layer pipelining, as long as the top-level inter-layer scheme satisfies the fine-grained forwarding requirements (e.g., with matched data forwarding granularities at the top level), we enable it when summing up the total pipeline runtime, i.e., waiting for one fmap instead of all (\cref{subsec:taxonomy} \ding{183}).

During cost estimation, \design{} adopts another pruning pass, skipping the schemes with non-Pareto-optimal access counts among the multiple tensors.

\textbf{Use of directives.}
To explore only the upper-level schemes, we target only the topmost \dir{GBUF} level directives without looking at the internal \tensor{}s and their \stack{} or \update{} operations.
The upper-level access counts, particularly DRAM and inter-node, are mostly determined by the top-level (\dir{GBUF}) \tensor{} and \update{} directives, and can be calculated as elaborated in \cref{subsec:directives}. Some of these directive parameters are constrained. For example, adjacent layers must have equal \tensor{} sizes and matched \update{} steps (see \cref{subsec:directives}). \design{} recognizes these constraints. 

\textbf{Dynamic-programming-based prioritization.}
We have now picked out a few potentially optimized inter-layer schemes for each segment, associated with their estimated costs. Next, to explore segment slicing and obtain the best overall segment chains for the entire NN, \design{} uses dynamic programming. It processes each layer in the DAG topological order, and in each step finds the segment chain that \emph{ends at} the current layer and has the minimum aggregated cost. As long as the cost estimation is accurate, dynamic programming provides optimized results in linear time. Nevertheless, to accommodate potential errors, instead of a single best segment chain, \design{} keeps the top $k_S$ candidates. We use a default $k_S$ as 4, and evaluate its impact in \cref{subsec:sensitivity}.

\subsection{Intra-Layer Stacking and Caching}
\label{subsec:solver-intra-layer}

With the short list of potentially optimized inter-layer schemes from \cref{subsec:solver-inter-layer}, \design{} continues to finalize the intra-layer scheduling. For a multi-level memory hierarchy, \design{} works through each level sequentially, from smaller, near-level buffers (e.g., \dir{REGF}) to larger, far-level ones (e.g., \dir{GBUF}). This bottom-up flow matches the common practice of dataflow optimization, which first maximally utilizes the faster and cheaper buffers to cache as many data as possible, and then gradually moves to slower and more expensive levels. This naturally ensures buffer capacity and parallelism constraints are \emph{always} satisfied during the exploration. In contrast, the traditional loop blocking approaches factorize the total loop factors across multiple buffer levels in a top-down manner. It cannot guarantee that the resultant tensors in each level fit in the buffer capacity, and thus produces a large number of invalid schemes and requires validity check on \emph{all} schemes.

Specifically, within each memory level, the key idea of \design{} is to maximize the \emph{sizes} of the tensors stored at this level, and to minimize the \emph{accesses} to the next level. We start from the smallest \emph{unit tensors}, with the same sizes as the tensors in the previous level. The unit tensor sizes at the first level are determined by the PE computation patterns. Then, we use two optimization passes, \emph{stacking} and \emph{caching}, to \emph{enlarge} the unit tensors along several carefully selected dimensions, until all the buffer capacity at this level is fully utilized, or the dimensions reach their total sizes and the tensors are fully stored in the buffers. Stacking parallelizes multiple tensors across buffers; caching increases the size of the tensor in each buffer. Both can reduce the accesses to the next level. We do stacking before caching, as stacking also improves parallelism.

The selected tensor dimensions to enlarge during stacking and caching must balance among the reuse of multiple tensors. This is a non-convex problem and hard to solve efficiently~\cite{blocking_cnn_arxiv16}. We use a greedy and iterative \emph{cost descending} method. At each step, we calculate the access count to each tensor, and choose a dimension that can help the tensor which has the maximum access count. We break ties using the second most accessed tensor. The chosen dimension is then slightly enlarged to its next smallest blocked size. We iterate multiple steps until the buffer capacity is used up. Thanks to our tensor-centric directives, both the necessary capacity checks and access count calculations can be done very fast.

\begin{figure}[t]
  \centering
  \includegraphics[width=\columnwidth]{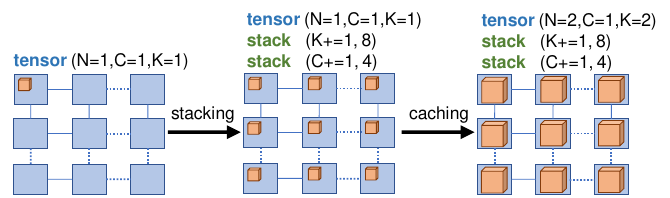}
  \caption{An example of intra-layer stacking and caching.}
  \label{fig:solver-intra-layer-example}
\end{figure}

\cref{fig:solver-intra-layer-example} shows an example. 
We use \stack{}s and enlarge \tensor{} dimension sizes for stacking and caching, respectively, from the inside out along the memory hierarchy. 
We can first decide the directive parameters at inner levels, and then freely select the dimensions to enlarge or stack at outer levels.
Initially, a unit tensor of size 1 is constructed. In the stacking pass, we select dimensions \dir{K} and \dir{C} to use $8 \times 4$ buffers, each storing a tensor with the original size of 1. The remaining factors along these dimensions are correspondingly reduced. Then in the caching pass, we select dimensions \dir{N} and \dir{K}, and enlarge the stored tensors to a size of \dir{N=2,C=1,K=2}. The next memory level will start from a unit tensor with this size, and work with the remaining dimension factors.

\input{algorithms/intra_layer_cost_descend}

\cref{algo:intra_layer_cost_descend} summarizes the intra-layer greedy cost descending algorithm.
We first initialize the unit tensor sizes and dataflow acquired by the specific hardware mapping (e.g., systolic array) and the NN layer. For each memory level bottom-up along the hierarchy, we iterate over different loop orders and apply stacking and caching schedules. Then we check various constraints (e.g., from inter-layer levels) and update the current best scheme at this memory hierarchy.

\newshrink{
Intra-layer scheduling must comply with various constraints. With layer pipelining, typically the tensor sizes at GBUF are fixed according to the data forwarding granularity~\cite{tangram_asplos19}. This is a challenge as \design{} starts from the innermost REGF, and cannot guarantee the final results at GBUF always satisfy the constraints. We instead first divide the total dimension sizes by these factors and directly schedule the blocked subtensors. If the result scheme completely fits these subtensors in GBUF, then the overall scheme satisfies the pipelining constraints.
}


%% file: algorithms/intra_layer_cost_descend.tex
\begin{algorithm}[tbh]
\caption{Greedy cost descending algorithm}
\label{algo:intra_layer_cost_descend}
\algnewcommand{\LineComment}[1]{\State \(\triangleright\) #1}
\begin{algorithmic}[1]
{\small
\State {\bf Input:} NN layer spec \texttt{LayerSpec}, hardware spec \texttt{HWSpec}, and constraints \texttt{Constrs}
\State {\bf Output:} Optimized dataflow scheme \texttt{OptDf}
\State \texttt{MemoryHier} = [\texttt{RegFile}, \texttt{GBUF}, \texttt{DRAM}]
\State \texttt{Df}, \texttt{LoopCnt} = \textsf{initUnitMapping}(\texttt{LayerSpec}, \texttt{HWSpec})
\For{each \texttt{m} in \texttt{MemoryHier}}
    \For{\texttt{order} \textbf{in} \textsf{genLoopOrder}(\texttt{m})}
        \State \texttt{Df}, \texttt{LoopCnt} = \textsf{stack}(\texttt{Df}, \texttt{m}, \texttt{LoopCnt}, \texttt{order})
        \State \texttt{Df}, \texttt{LoopCnt} = \ \textsf{cache}(\texttt{Df}, \texttt{m}, \texttt{LoopCnt}, \texttt{order})
        \If{\textsf{checkValid}(\texttt{Constrs}, \texttt{LoopCnt}) is \texttt{False}}
            \State \textbf{continue}
        \EndIf
        \If{\textsf{evalCost}(\texttt{Df}) $>$ \textsf{evalCost}(\texttt{OptDf})}
            \State \texttt{OptDf} = \texttt{Df}
        \EndIf
    \EndFor
\EndFor
}
\end{algorithmic}
\end{algorithm}

%% file: 5_methodology.tex
\section{Methodology}
\label{sec:method}

{\bf Evaluated NNs.}
We evaluate both inference and training on five CNNs, AlexNet~\cite{alexnet_nips12}, MobileNet~\cite{mobilenets_arxiv17}, VGGNet~\cite{vggnet_arxiv14}, GoogLeNet~\cite{googlenet_cvpr15}, and ResNet~\cite{resnet_cvpr16}, as well as an MLP~\cite{prime_isca16} and an LSTM~\cite{lstm_gnmt_nips14}.
We use a default batch size of 64, but for inference on small edge devices we use batch size 1.

{\bf Hardware architectures.}
Our target architecture (\cref{fig:multi_node_arch}) consists of $16 \times 16$ nodes, each of which contains an $8 \times 8$ PE array with a \bytes{64} register file per PE, and a \kb{32} global buffer~\cite{tangram_asplos19}. In total there are 16384 PEs and \mb{8} on-chip SRAM. We model it in \SI{28}{\nano\meter} and assume a \SI{500}{\mega\hertz} logic frequency. The PE array uses Eyeriss-like row-stationary mapping~\cite{eyeriss_isca16}. The 16-bit MAC in a PE costs \SI{1}{\pico\joule}~\cite{eyeriss_isscc16, shidiannao_isca15, flexflow_hpca17}. McPAT~1.3 models the register files and SRAM buffers of different sizes, as well as the PE array buses~\cite{mcpat_micro09}. The NoC energy is \SI{0.61}{\pico\joule\per bit} per each hop~\cite{jenga_isca17}. The off-chip memory supplies \SI{25.6}{\giga Bps} bandwidth with four LPDDR4 channels, with energy modeled from commercial datasheets~\cite{micron_lpddr4}.

Besides the large multi-node accelerator, we also evaluate a small single-node edge inference device. It contains a $16 \times 16$ PE array with \bytes{512} registers per PE, and a \kb{256} global buffer. We intentionally use a different, TPU-like systolic array~\cite{tpu_isca17} to demonstrate the generality of \design{}.

With the lack of actual hardware, we use the \texttt{nn-dataflow} simulator~\cite{tetris_asplos17, tangram_asplos19} to model the above two architectures. 
This simulator is validated against both cycle-accurate simulation and real Eyeriss results~\cite{tetris_asplos17}. It has high accuracy by considering both compute time and data transfer time at the minimum granularity of blocked data for one round of PE array execution.
We extend it to support training and systolic arrays. All reported performance and energy consumption results are from this simulator. Notice that this is a different, much more detailed and accurate cost model compared to that in \design{}. The \design{} cost model is only used to guide the solving process, and does not provide precise evaluation.

{\bf Baseline solvers.}
We compare \design{} (\texttt{K}) with state-of-the-art approaches. The baseline (\texttt{B}) is \texttt{nn-dataflow}~\cite{tangram_asplos19} which uses exhaustive search to find the globally optimal schemes. Its search speed is highly optimized with parallel execution and pruning. We implement another exhaustive search (\texttt{S}) using our directives. The random search (\texttt{R}) from Timeloop~\cite{timeloop_ispass19} evaluates candidates at each level with a given probability, except for segment slicing (skipping segments may not result in complete segment chains). We empirically find the probability should be no less than 0.1 at each level to guarantee finding valid schemes. The ML-based method (\texttt{M}) from AutoTVM~\cite{autotvm_neurips18} uses simulated annealing guided by XGBoost to handle intra-layer scheduling, while exhaustively exploring inter-layer options. For each layer we run 1024 iterations that each tunes a batch of 128 configurations.

%% file: 6_evaluation.tex
\section{Evaluation}
\label{sec:eval}

\subsection{Dataflow Efficiency and Scheduling Speed}
\label{subsec:results}

\begin{figure}
  \centering
  \includegraphics[width=\columnwidth]{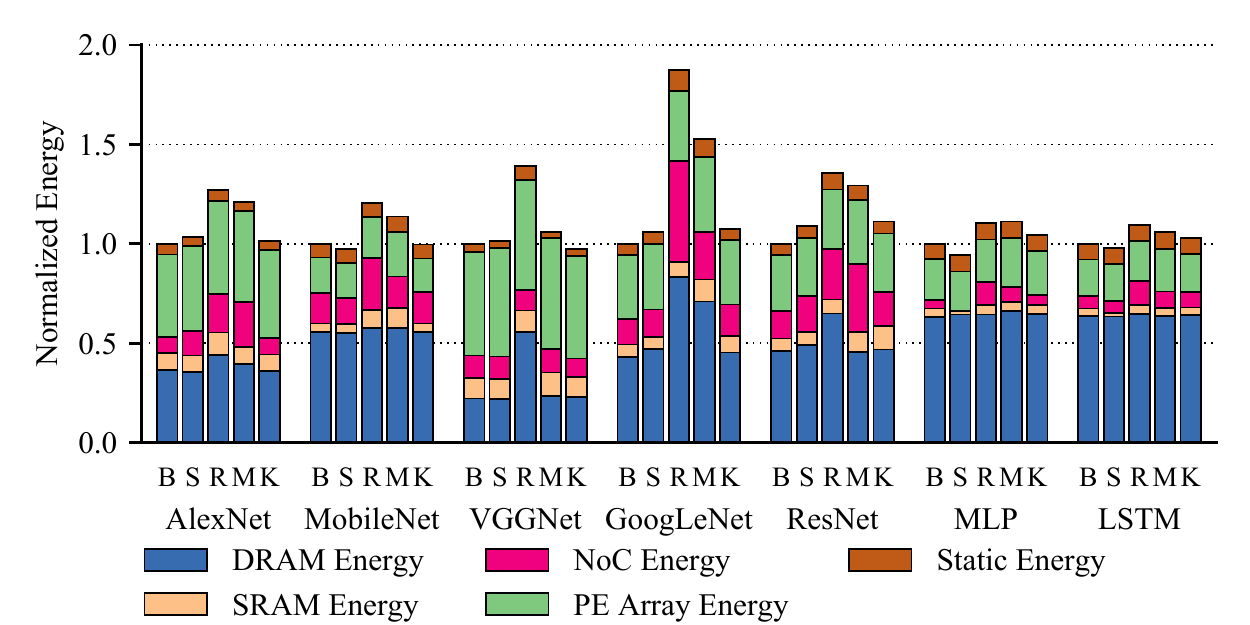}
  \caption{Comparison of dataflow energy for training on multi-node Eyeriss-like accelerators with batch 64.}
  \label{fig:training_energy}
\end{figure}

For training on the large multi-node architecture, \cref{fig:training_energy} illustrates the energy of the resulted dataflow schedules from the five scheduling approaches, normalized to the baseline \texttt{B}. First, exhaustively searching on our tensor-centric directives (\texttt{S}) is able to match the baseline efficiency, demonstrating the generality of the directives. In some cases like MobileNet and MLP, \texttt{S} even results in slightly better schemes, because the directives are more flexible and support a larger design space than prior work. Second, random search \texttt{R} is the least robust method, whose results vary significantly across different NNs, with up to 1.9$\times$ worse energy. This demonstrates that randomly skipping candidates without considering the design space characteristics is insufficient. Third, the ML-based \texttt{M} works better than \texttt{R}, with 17.7\% on average and up to 1.5$\times$ worse results than optimal. We suspect that complex NNs like GoogLeNet and ResNet have larger design spaces and more constraints, which are hard for ML algorithms to learn.

Compared to the exhaustively searched optimal, \design{} only has a 2.2\% energy overhead on average, and within 10\% for the worst case MLP. \shrink{This demonstrates \design{}'s effectiveness. }
Furthermore, the energy breakdowns across major hardware components also match well. As different components are mainly affected by different scheduling levels, the results indicate that the design space decoupling of \design{} works well and each level's solver is individually optimized.

\begin{table}
    \centering\tablefontsize
    \caption{Comparison of scheduling time for NN training on multi-node accelerators. Measured on an Intel Xeon Gold 5120 processor with 8 parallel processes.}
    \label{table:bp_scheduling_times}
    \newcommand{\thr}[1]{\SI{#1}{\hour}}
    \newcommand{\tmi}[1]{\SI{#1}{\minute}}
    \newcommand{\tse}[1]{\SI{#1}{\second}}
    \begin{threeparttable}
    \begin{tabular}{lrrrrr}
        \toprule
        & \makecell[c]{\texttt{B}} & \makecell[c]{\texttt{S}} & \makecell[c]{\texttt{R}} & \makecell[c]{\texttt{M}} & \makecell[c]{\texttt{K}} \\
        \midrule
        {\bf AlexNet} & \thr{8.7} & \thr{12.7} & \tmi{4.6} & \thr{3.2} & \tse{32} \\
        {\bf MobileNet} & \thr{17.6} & \thr{33.8} & \tmi{16.6} & \thr{10.1} & \tse{53} \\
        {\bf VGGNet} & \thr{6.8} & \thr{12.2} & \tmi{5.5} & \thr{3.6} & \tse{35} \\
        {\bf GoogLeNet} & \thr{64.2} & \thr{111.0} & \tmi{31.4} & \thr{37.1} & \tse{262} \\
        {\bf ResNet} & \thr{78.6} & \thr{114.8} & \tmi{42.4} & \thr{28.6} & \tse{118} \\
        {\bf MLP} & \thr{0.2} & \thr{0.7} & \tmi{1.0} & \thr{0.2} & \tse{13} \\
        {\bf LSTM} & \thr{0.2} & \thr{0.9} & \tmi{7.4} & \thr{0.1} & \tse{6} \\
        \bottomrule
    \end{tabular}
    \end{threeparttable}
\end{table}

For scheduling speed in \cref{table:bp_scheduling_times}, \design{} significantly outperforms all alternative approaches. Compared to the baseline \texttt{B}, \design{} offers 518$\times$ speedup on average. It is also 265$\times$ faster than ML-based \texttt{M}, as the ML-based method needs to iterate many steps to converge. The more complex an NN is, the more speedup \design{} can achieve. While the other scheduling methods spend hours or days, \design{} can typically output optimized schemes within a few seconds and up to a few minutes. In summary, \design{} realizes a more practical approach towards real-time interactive compilation of optimized dataflow on large-scale NN accelerators.

\begin{figure}
    \centering
    \includegraphics[width=\columnwidth]{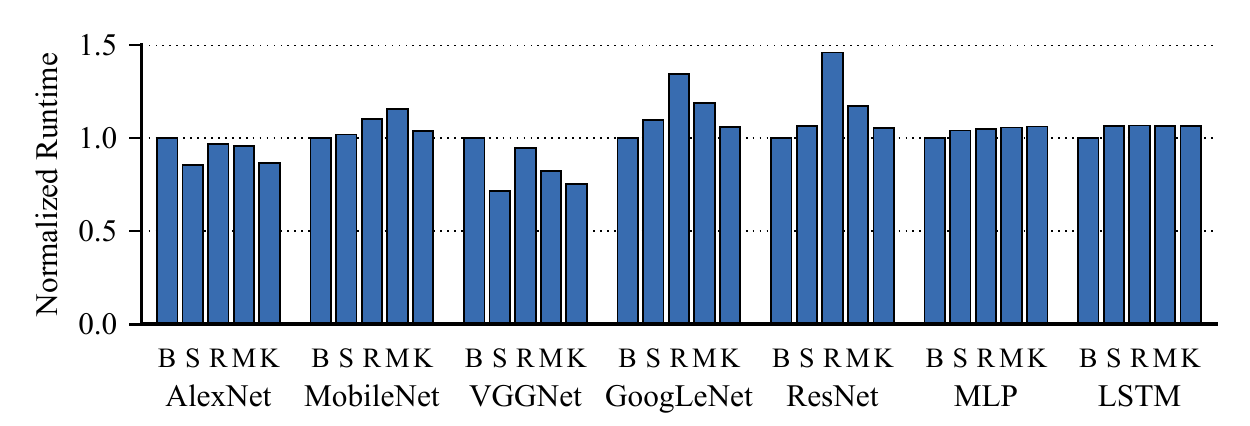}
    \caption{Comparison of dataflow performance for training on multi-node Eyeriss-like accelerators with batch 64.}
    \label{fig:training_performance}
\end{figure}

\design{} can also be used to optimize the performance as introduced in \cref{sec:solver}. \cref{fig:training_performance} shows that the performance results generally follow the same trends.
This validates our conjecture of co-optimizing energy and performance.

\begin{figure}
    \centering
    \includegraphics[width=\columnwidth]{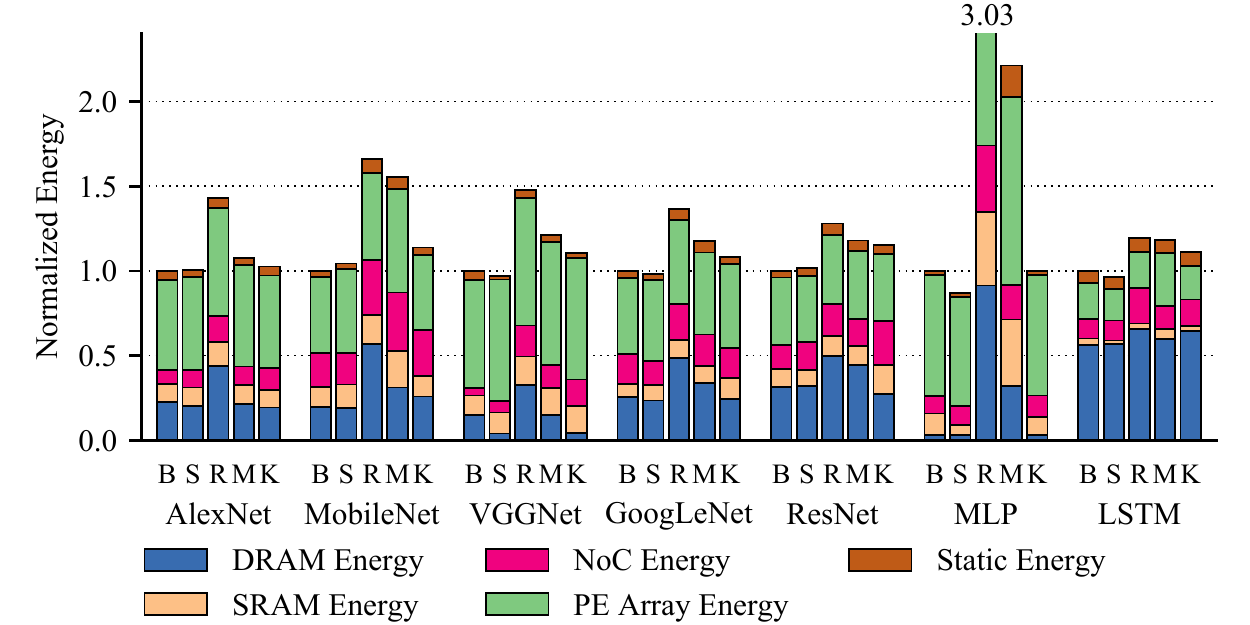}
    \caption{Comparison of dataflow energy for inference on multi-node Eyeriss-like accelerators with batch 64.}
    \label{fig:inference_energy}
\end{figure}

We also evaluate inference on the multi-node architecture in \cref{fig:inference_energy}. Without back-propagation layers, the inference DAGs are much simpler than the training DAGs and have fewer constraints~\cite{tangram_asplos19}. Therefore the scheduling becomes more involved with more options. Nevertheless, \design{} has only a 7.7\% energy overhead on average compared to \texttt{B}, but is 174$\times$ faster. It is also 50$\times$ and 1.5$\times$ faster than \texttt{M} and \texttt{R}, while the latter two have much worse average efficiency drops of 36.1\% and 59\%, respectively. Notice that the performances of ML-based and random-based methods are poor on MLP. We attribute this to the low computation-storage ratio of MLP, which further leads to tight storage constraints and hinders the discovery of good or even valid dataflows of the two methods.



\begin{figure}
  \centering
  \includegraphics[width=\columnwidth]{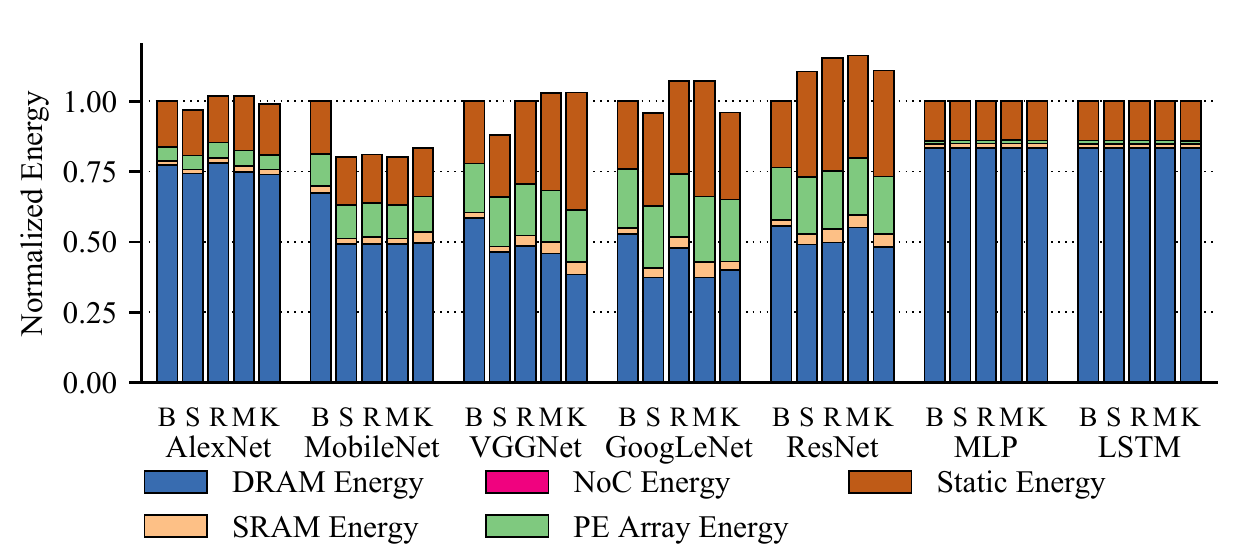}
  \caption{Comparison of dataflow energy for inference on single-node TPU-like accelerators with batch 1.}
  \label{fig:systolic_energy}
\end{figure}

\cref{fig:systolic_energy} further demonstrates \design{}'s generality on a small edge accelerator with only \kb{256} buffer, and a TPU-like systolic array. Since the design space is small, all methods can achieve near-optimal efficiency. The ML-based method has an overhead up to 16\%. We have to set the random search probability to 0.85 to get valid schemes under the rigid constraints of a small on-chip buffer, which leads to an average of 3.8\% energy overhead. This indicates that ML- and random-based approaches suffer from generality issues in setting their hyperparameters. \design{} introduces a 1.9\% overhead on average, with the worst case 10\% on ResNet.
These results show that \design{} is robust even in resource-limited edge scenarios.

\subsection{Sensitivity Studies and Model Validation}
\label{subsec:sensitivity}

\begin{table}
    \centering\tablefontsize
    \caption{Energy overheads of the \design{} dataflow results for GoogLeNet with different hardware configurations.}
    \label{table:hardware_config_sweep}
    \begin{threeparttable}
    \begin{tabular}{cccccc}
        \toprule
        {\bf Batch} & {\bf Nodes} & {\bf PEs} & {\bf GBUF} & {\bf REGF} & {\bf Overhead} \\
        \midrule
        64  & $4 \times 4$ & $8 \times 8$     & \kb{32}  & \bytes{32}  & 4.4\% \\
        64  & $4 \times 4$ & $8 \times 8$     & \kb{32}  & \bytes{64}  & 2.7\% \\
        64  & $4 \times 4$ & $8 \times 8$     & \kb{32}  & \bytes{128} & 1.5\% \\
        8   & $4 \times 4$ & $16 \times 16$   & \kb{32}  & \bytes{32}  & 8.3\% \\
        1   & $16 \times 16$ & $8 \times 8$   & \kb{32}  & \bytes{64}  & 6.7\% \\
        \bottomrule
    \end{tabular}
    \end{threeparttable}
\end{table}

\cref{fig:systolic_energy} shows \design{} works with different PE array architectures.
\cref{table:hardware_config_sweep} further shows how \design behaves on more hardware configurations.
The overheads are generally small, demonstrating that \design{} is robust to hardware changes.


\begin{figure}
  \centering
  \includegraphics[width=\columnwidth]{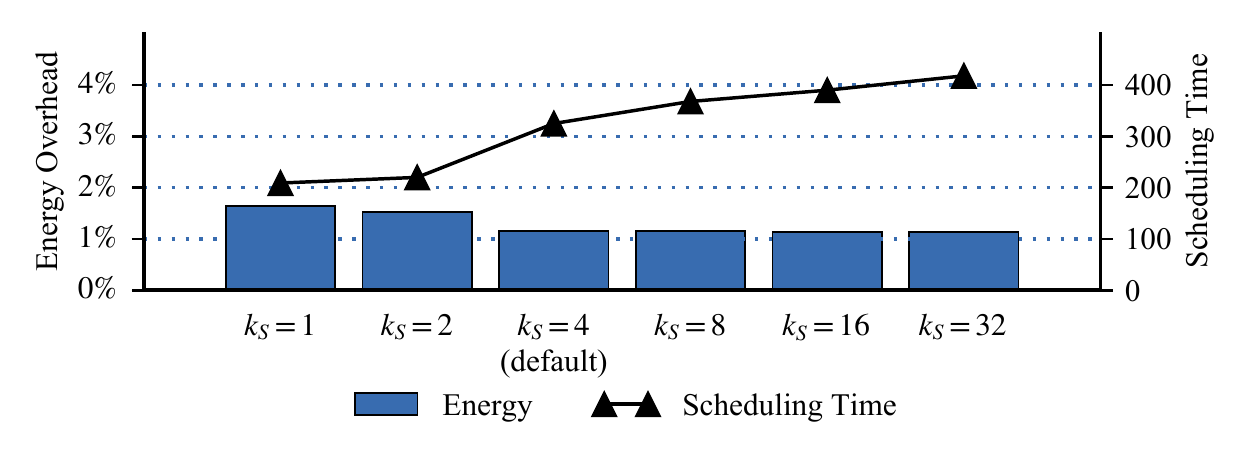}
  \caption{Impact of segment candidate counts in \design{}.}
  \label{fig:seg_hyper_para}
\end{figure}

\design{}'s dynamic-programming-based prioritization tolerates cost estimation errors by keeping up to $k_S$ candidate segments. The default value is 4. We study its impact on the tradeoff between dataflow efficiency and scheduling time, as shown in \cref{fig:seg_hyper_para}.
We see that as $k_S$ decreases, the energy overheads do not increase much, meaning that the cost estimation errors are insignificant. However the search speed substantially improves.

\begin{table}
    \centering\tablefontsize
    \caption{Effectiveness of inter-layer conservative pruning. Only one representative segment is reported for each NN.}
    \label{table:pruning_ablation_study}
    \begin{threeparttable}
    \begin{tabular}{lccc}
        \toprule
        & {\bf Total Schemes} & {\bf After Pruning} & {\bf \% Pruned} \\
        \midrule
        {\bf AlexNet} & 700 & 2 & 99.7\% \\
        {\bf MobileNet} & 331 & 42 & 87.3\% \\
        {\bf VGGNet} & 7 & 1 & 85.7\% \\
        {\bf GoogLeNet} & 560 & 1 & 99.8\% \\
        {\bf ResNet} & 63 & 4 & 93.6\% \\
        {\bf MLP} & 784 & 18 & 97.7\% \\
        {\bf LSTM} & 70 & 3 & 95.7\% \\
        \bottomrule
    \end{tabular}
    \end{threeparttable}
\end{table}

\cref{table:pruning_ablation_study} shows the effectiveness of \design{}'s pruning techniques used in the inter-layer phase, including both validity and Pareto-optimal pruning (\cref{subsec:solver-inter-layer}). For each NN, we pick one segment to count its schemes before and after pruning. As we can see there could be hundreds of inter-layer dataflow schemes for \emph{each} of the tens to hundreds of segments, each with a large number of intra-layer options to be explored, making the design space quite huge. Pruning successfully eliminates most invalid or suboptimal choices, speeding up the solver without sacrificing accuracy.

\shrink{
The effectiveness of \design{} relies on its ability to accurately estimate dataflow cost from our tensor-centric directives. We compare this cost estimation against the validated baseline \texttt{nn-dataflow} model, and see that the estimation successfully matches the accurate cost model, with deviations no more than 11\%.
}

%% file: 7_related_work.tex
\section{Related Work}

There has been a large body of work on NN acceleration, either designed as domain-specific chips~\cite{neuflow_cvprw11, diannao_asplos14, dadiannao_micro14, shidiannao_isca15, eyeriss_isscc16, neurocube_isca16, cambricon_isca16, flexflow_hpca17, tetris_asplos17, tpu_isca17, scaledeep_isca17} or prototyped on reconfigurable FPGAs~\cite{cnn_fpga_iccd13, cnn_fpga15, caffeine_cnn_fpga_iccad16, fused_layer_cnn_micro16, cnn_pipeline_fpl16, dnn2fpga_micro16, cnn_fpga_util_isca17, brainwave_dnn_fpga_isca18}. This diverse set of hardware results in a rich set of dataflow choices.
The taxonomy proposed in our work is sufficiently expressive to describe these dataflow schemes. See \cref{table:taxonomy} for a summary.

To formally represent dataflow schemes, Timeloop~\cite{timeloop_ispass19} used nested loops. MAESTRO~\cite{maestro_micro19} transformed nested loops into directives such as \texttt{TemporalMap} and \texttt{SpatialMap}, to better expose data reuse opportunities. Interstellar~\cite{interstellar_asplos20} borrowed scheduling primitives from Halide~\cite{halide_pldi13}, which are also designed for nested-loop-based programs. There also exist high-level domain-specific languages and IRs for NN compilation on CPUs/GPUs~\cite{mlir_arxiv20, glow_arxiv18}. As said in \cref{subsec:challenges}, these frameworks are neither directly applicable to the comprehensive design space of multi-node NN accelerators, nor pragmatic to guide their dataflow scheduling.

In terms of NN dataflow search methods, Yang et al.~\cite{blocking_cnn_arxiv16} were among the first to comprehensively describe the rich design space of loop transformations and show the difficulty of general scheduling. Timeloop~\cite{timeloop_ispass19} introduced a cost model and a scheduling mapper, but still relied on exhaustive or random search. HyPar~\cite{hypar_hpca19} and AccPar~\cite{accpar_hpca20} explored intra-layer parallelism using more effective algorithms such as dynamic programming, but did not study detailed loop transformations or segment slicing. Targeting CPUs/GPUs, Jia et al.~\cite{nn_para_gpu_icml18, nn_para_gpu_sysml19} summarized different layer parallelization strategies and proposed an optimization algorithm. Mirhoseini et al.~\cite{device_placement_rl_icml2017, device_placement_rl_iclr2018} adopted reinforcement learning to schedule NNs onto heterogeneous CPU-GPU clusters. To generate optimized GPU code for individual layers, AutoTVM~\cite{autotvm_neurips18} used simulated annealing combined with XGBoost, and TensorComprehensions~\cite{tensor_comprehensions_arxiv18} and GAMMA~\cite{gamma_iccad20} used genetic algorithms. Ahn et al.~\cite{device_placement_rl_iclr2018} formulated the exploration as a reinforcement learning problem. MAGNet~\cite{magnet_iccad19} used Bayesian optimization to search both the architecture and the dataflow.  MindMappings~\cite{mindmappings_asplos21} used an NN to approximate the design space of intra-layer and single-node architectures.


%% file: 8_conclusion.tex
\section{Conclusions}

In this paper we present a hierarchical taxonomy and a set of tensor-centric directives to describe the rich NN dataflow space.
We then build a generic, optimized, and fast dataflow solver, \design{}, and demonstrate its effectiveness when scheduling complex NNs on various scales of accelerators for training and inference. \design{} achieves near-optimal schedules, and is significantly faster than previous methods.